\documentclass[
 reprint,
 nofootinbib,
 amsmath,amssymb,
 aps,
 prd,
 floatfix,
]{revtex4-1}

\usepackage{graphicx}
\usepackage{bm}
\usepackage{booktabs}
\usepackage{dcolumn}
\usepackage{xcolor}
\usepackage{hyperref}
\usepackage{xspace}
\hypersetup{colorlinks=true,citecolor=blue,urlcolor=blue,linkcolor=blue}
\usepackage[normalem]{ulem}

\newcommand{\wb}{\omega_b}
\newcommand{\fede}{f_{\rm vEDE}}
\newcommand{\zc}{z_c}
\newcommand{\dH}{\Delta H/H}
\newcommand{\dNeff}{\Delta N_{\rm eff}}
\newcommand{\primat}{\texttt{PRIMAT}\xspace}
\newcommand{\parthenope}{\texttt{PArthENoPE}\xspace}
\newcommand{\TD}{T_\mathrm{D}}
\newcommand{\dD}{\delta_\mathrm{D}}

\usepackage{tikz,xcolor}

\definecolor{lime}{HTML}{A6CE39}
\DeclareRobustCommand{\orcidicon}{\hspace{-4pt}
	\begin{tikzpicture}
		\draw[lime, fill=lime] (0,0) 
		circle [radius=0.16] 
		node[white] {\hspace{0.1mm}{\fontfamily{qag}\selectfont \tiny ID}};
		\draw[white, fill=white] (-0.07,0.1) 
		circle [radius=0.01];
	\end{tikzpicture}
	\hspace{-3.2mm}
}

\foreach \x in {A, ..., Z}{\expandafter\xdef\csname 
	orcid\x\endcsname{\noexpand\href{https://orcid.org/\csname orcidauthor\x\endcsname}
		{\noexpand\orcidicon}}
}

\begin{document}

\title{What could an emerging Big Bang Nucleosynthesis discrepancy be hinting at?}

\author{Vivian Poulin\orcidV{}}
\email{vivian.poulin@umontpellier.fr} 
\affiliation{Laboratoire Univers \& Particules de Montpellier (LUPM), CNRS \& Universit\'{e} de Montpellier (UMR-5299), Place Eug\`{e}ne Bataillon, F-34095 Montpellier Cedex 05, France}

\author{Julien Froustey\orcidJ{}}
\email{julien.froustey@ific.uv.es}
\affiliation{Institut de Física Corpuscular (IFIC), CSIC-Universitat de València, Parc Científic UV, C/ Catedrático José Beltrán 2, E-46980 Paterna (València), Spain}

\author{Cyril Pitrou\orcidC{}}
\email{cyril.pitrou@iap.fr}
\affiliation{Institut d’Astrophysique de Paris, Sorbonne Université, CNRS UMR 7095, 98 bis bd Arago, 75014 Paris, France}

\author{Tristan L. Smith\orcidT{}}
\email{tsmith2@swarthmore.edu}
\affiliation{Department of Physics and Astronomy, Swarthmore College, 500 College Ave., Swarthmore, PA 19081, USA}

\date{\today}

\begin{abstract}
The latest measurement of the primordial deuterium abundance is in $\sim 2\sigma$ tension with several state-of-the-art predictions of standard big bang nucleosynthesis (BBN), when using the baryon density inferred from the $\Lambda$CDM model fit to cosmic microwave background (CMB) data. This tension increases to $\sim 3\sigma$ for models attempting to solve the Hubble tension, such as early dark energy (EDE), which generally predict a larger baryon density than in $\Lambda$CDM.  
We test whether this discrepancy could be pointing to a non-standard
expansion history during BBN.  We compute light-element
abundances with \primat and compare $\Lambda$CDM, a $\dNeff$ extension, and a very early dark energy (vEDE) component.  
For vEDE, we sample $\Delta H/H$, the fractional increase of the expansion rate while deuterium burning is freezing out and helium-4 fusion is mostly over. 
The Bayesian analysis using BBN plus the CMB baryon-density constraint in the EDE cosmology
 gives $\Delta H/H = 0.087^{+0.036}_{-0.037}$ during the deuterium burning epoch (i.e., at a temperature $\TD \simeq0.03\,{\rm MeV}$), and no residual tension.  The vEDE component preserves the observed deuterium abundance at the larger CMB baryon density while only mildly affecting helium-4.  By contrast, extra
radiation raises the helium-4 abundance too efficiently and does not reconcile the baryon density determinations. Together with inflation, dark energy, and EDE, our results hint at the presence of another light scalar-field in cosmology.
\end{abstract}

\maketitle

\section{Introduction}
\label{sec:intro}

Big Bang Nucleosynthesis (BBN) is a precision probe of the expansion
history at temperatures well above those directly tested by the cosmic
microwave background (CMB).  At present, however, modern BBN calculations do
not give identical predictions for the primordial deuterium abundance (D/H), even
though they agree well on the helium-4 prediction~\cite{Pitrou:2018cgg,Gariazzo:2021iiu,Yeh:2020mgl,Pisanti:2020efz,Pitrou:2020etb,Burns:2023sgx,Launders:2026ciu}.  The difference between
BBN codes mainly traces back to the selection and treatment of the deuterium-burning nuclear
rates, which require converting sparse
low-energy data into thermonuclear rates, and different defensible choices
shift the predicted residual D/H~\cite{Pitrou:2021vqr}.  In particular, the most up-to-date implementation of the BBN code \primat, together with its adopted nuclear reaction rates,
leads to a mild discrepancy with the CMB baryon density inferred in
$\Lambda$CDM.  Whether this discrepancy ultimately survives improved nuclear
input and abundance measurements remains to be
seen, but it is already useful to ask what it would be telling us if it
persisted or grew.

BBN is sensitive to the baryon density $\wb\equiv\Omega_b h^2$ through the baryon-to-photon ratio $\eta$
(see, e.g.,~\cite{Steigman:2006nf} for the relation between these quantities). The baryon density controls in particular the efficiency of deuterium burning, with
${\rm D/H}\propto\wb^{-1.65}$ around the $\Lambda$CDM solution. 
The primordial deuterium abundance is measured at percent precision
using metal-poor absorption systems. Using the Cooke \emph{et al.}
measurement~\cite{Cooke:2018}
\begin{equation}
  10^5\,{\rm D/H}=2.527\pm0.030,
  \label{eq:dhobs}
\end{equation}
In standard BBN in the \primat setup, this D/H measurement alone gives
approximately
\begin{equation}
\label{eq:wb_DHonly}
 \wb^{\rm BBN} \simeq 0.0219 \pm 0.00022
\end{equation}
where the error bar include both observational and \primat nuclear rates uncertainty.\footnote{More recent
measurements and compilations recommend the slightly lower value
$10^5\,{\rm D/H}=2.508\pm0.030$
\cite{Kislitsyn:2024jvk,Schoneberg:2024ifp,ParticleDataGroup:2026}.  In the
same \primat setup, this lower D/H value would shift the inferred baryon
density upward to $\wb\simeq0.0220$, reducing the tension to about
$1.6\sigma$ with the $\Lambda$CDM CMB value quoted below, or about
$2.0\sigma$ with the higher CMB+DESI $\Lambda$CDM value.  We retain
Eq.~\eqref{eq:dhobs} in the main analysis because it has been widely used in
recent BBN comparisons, and present the lower-D/H results in
Appendix~\ref{app:dh2508}.} This lies roughly $2.1\sigma$
below the $\Lambda$CDM CMB value
$\wb=0.022398\pm0.000095$ inferred from the Planck NPIPE
\cite{Rosenberg:2022sdy}, ACT~\cite{Louis:2025act}, and
SPT~\cite{Camphuis:2025spt} CMB data combination, hereafter SPA
\cite{Camphuis:2025spt}.  Even within $\Lambda$CDM, the
inclusion of DESI BAO data pulls the CMB-inferred baryon density upward; for
example the corresponding CMB+DESI value is
$\wb=0.022478\pm0.000091$~\cite{Camphuis:2025spt}, increasing the tension to
$2.4\sigma$.

If this discrepancy persists, modified expansion during BBN is a
natural phenomenological possibility to explore.  Once the deuterium
bottleneck opens, nuclear reactions have only a finite time to process
deuterium into heavier nuclei.  Increasing the expansion rate shortens this
interval and raises the surviving D/H abundance
\cite{Froustey:2019owm,1979ApJ...227..697Y,1984ApJ...281..493Y}. Consequently, the same value of D/H is obtained with a larger $\omega_b$, thus removing the tension. The key
question is not only whether such a change can accommodate D/H, but whether it
can do so without spoiling the success of helium-4.  The helium-4 abundance is
well measured, and its theoretical prediction is comparatively robust across
modern BBN codes because it is controlled mainly by the neutron-to-proton ratio
at weak freeze-out rather than by the late nuclear rates.

The recent Large Binocular Telescope (LBT) determination of the
helium-4 abundance~\cite{Aver:2026}
\begin{equation}
  Y_p=0.2458\pm0.0013 \, ,
  \label{eq:ypobs}
\end{equation}
provides the complementary weak-freeze-out probe used in this work
\cite{Yeh:2026pil}.  Measurements on extremely metal-poor galaxies by the
EMPRESS collaboration have resulted in a lower value for $Y_p$
\cite{Matsumoto:2022tlr,Yanagisawa:2025mgx}.  In this work, we retain the most
recent determination~\eqref{eq:ypobs}, noting that a lower $Y_p$ could be
accommodated, for instance, by a nonzero electron neutrino chemical
potential~\cite{Burns:2022hkq,Escudero:2022okz,Froustey:2024mgf}.
This
helium constraint is precisely why a constant radiation excess,
parameterized by $\Delta N_{\rm eff}$, is already tightly constrained by BBN:
the same extra radiation that raises D/H also changes weak freeze-out and
therefore $Y_p$ \cite{Steigman:2007xt,Cyburt:2015mya,Fields:2019pfx,Pitrou:2018cgg}.
It is therefore natural to ask whether a transient expansion episode can
instead increase $H$ during the deuterium-burning epoch while leaving weak
freeze-out and helium-4 production largely unaffected.

Early dark energy (EDE) provides a concrete setting in which to test this idea. EDE has
emerged as a promising framework to address the Hubble tension between direct
measurements of the Hubble constant and the value inferred from the CMB
assuming $\Lambda$CDM
\cite{Poulin:2018cxd,Schoneberg:2021qvd,Kamionkowski:2022pkx,Poulin:2025nfb}.
By increasing the pre-recombination expansion rate, EDE reduces the sound
horizon and allows for a larger CMB-inferred value of $H_0$.  Yet this is
accompanied by a higher baryon density inferred from the CMB power spectrum:
the latest EDE analysis combining up-to-date CMB, BAO, and
SH0ES-calibrated Pantheon+ SN1a data yields
\cite{Schoneberg:2026eys,Schoneberg:2026vaf}
\begin{equation}
\label{eq:wb_EDE}
 \wb = 0.02272\pm 0.00014\,,
\end{equation}
which is roughly $3.1\sigma$ above the \primat D/H-only value in
Eq.~\eqref{eq:wb_DHonly}.  More generally, a positive $H_0$--$\wb$ correlation
is expected in pre-recombination solutions to the Hubble tension~\cite{Giovanetti:2026aku}.  This behavior is found in the recent ``$H_0$
World Cup'' analyses~\cite{Schoneberg:2026eys,Schoneberg:2026vaf} (see Figs.~21 \& 22 in~\cite{Schoneberg:2026vaf})---a comprehensive benchmarking of models proposed to resolve the Hubble tension--across the successful early-universe models, whether
they alter recombination physics itself or the pre-recombination expansion history, as in EDE.
Therefore, if EDE or a related early-time solution
ultimately proves correct, the higher CMB baryon density would exacerbate the
BBN discrepancy and make the case for additional BBN-era physics even
sharper~\cite{Giovanetti:2026aku}.

The ``$\omega_b$ tension'' is not identical in all modern BBN
calculations: the standard D/H prediction of \parthenope
\cite{Pisanti:2020efz,Gariazzo:2021iiu}, or that of Ref.~\cite{Yeh:2020mgl},
is in closer agreement with Eq.~\eqref{eq:dhobs}.  The difference mainly
comes from the deuterium-burning nuclear rates.  While
$\mathrm{D}(p,\gamma){}^3\mathrm{He}$ has been precisely measured
\cite{Mossa:2020gjc}, the $\mathrm{D}(\mathrm{D},n){}^3\mathrm{He}$ and
$\mathrm{D}(\mathrm{D},p){}^3\mathrm{H}$ rates remain less certain, and
different justified conversions of the data into nuclear rates lead to
different predictions in \primat and \parthenope~\cite{Pitrou:2021vqr}.
Other codes, such as \texttt{PRyMordial}~\cite{Burns:2023sgx}, allow the user
to choose between rate sets.  A recent independent approach using
\texttt{LINX}~\cite{Giovanetti:2024zce} and Gaussian-process nuclear-rate
fits also hints at a mild baryon-density tension, with results close to the
\primat ones~\cite{Launders:2026ciu}. 
Note, however, that even the \parthenope determination of $\omega_b$ is in
small tension with early-universe solutions that prefer a high CMB baryon
density~\cite{Schoneberg:2026eys,Schoneberg:2026vaf}.

In this work, we use a new Python implementation of \primat to ask
whether the emerging BBN--CMB baryon-density discrepancy could be a sign of a
non-standard expansion history during BBN, and which kind of modification is
required.  We compare $\Lambda$CDM, a constant radiation contribution
parameterized by $\dNeff$, and a model involving an additional transient
\emph{very} early dark energy (vEDE) contribution during BBN.  Related studies
have constrained specific transitions during BBN and model-independent
modifications of the expansion history~\cite{McKeen:2024jve,Cook:2025gra}.

From this perspective, vEDE is not introduced only as a fix for a
specific cosmological model, but as one possible explanation of what a
persistent BBN discrepancy could be hinting at.  Cosmology already entertains
several epochs in which scalar fields may dominate or significantly affect the
expansion history, including inflation, late-time dark energy, and perhaps
EDE.  It is therefore natural to ask whether another transient episode during
BBN could account for the discrepancy.  For this reason, our main combined BBN+CMB analyses use
the high baryon density inferred in the EDE cosmology as a representative
target: it makes the BBN discrepancy sharper and connects the possible
BBN-era episode to the broader sequence of scalar-field epochs.  We also
repeat the analysis using the lower $\Lambda$CDM CMB baryon-density
constraint in Appendix~\ref{app:lcdmprior}. 

Our main result below is that an increase in expansion rate can resolve this emerging tension, and the timing of the expansion-rate enhancement is
essential: a transient vEDE component within the deuterium-burning epoch can preserve deuterium at large $\wb$
without producing the helium shift associated with constant extra radiation. This results in an upper limit to the time at which 
the vEDE field becomes dynamical, or equivalently, an upper limit on the effective 
mass of the field.

The cosmological consequences of such very early transient components
are not limited to the homogeneous expansion history.  Ref.~\cite{Sobotka:2024tat}
studied for the first time scalar-field vEDE models that become dynamical after BBN and before
matter-radiation equality, showing that they can leave distinctive signatures
in the small-scale matter power spectrum while producing relatively small changes in
the primary CMB when the transition occurs sufficiently early.  In particular,
if the vEDE component temporarily dominates the energy density, modes entering
the horizon near the transition can develop a bump in power, whereas modes on
neighboring scales are suppressed.  Our analysis is complementary: rather than
using late-time structure to search for these small-scale signatures, we ask
whether a vEDE contribution during BBN can reconcile the light-element and CMB
baryon-density determinations.

The rest of the paper is structured as follows.  We introduce the models and
motivate our vEDE parameterization in Sec.~\ref{sec:models}.  The \primat
calculation, datasets, and statistical analysis are described in
Sec.~\ref{sec:method}.  We present the PL and MCMC results in
Sec.~\ref{sec:results}, before discussing their implications and concluding
in Sec.~\ref{sec:conclusion}.

\section{Models}
\label{sec:models}

\subsection{\texorpdfstring{$\Lambda$CDM}{LambdaCDM} and
\texorpdfstring{$\dNeff$}{Delta Neff}}

Our baseline BBN calculation assumes the $\Lambda$CDM expansion history and
the Standard Model prediction $N_{\rm eff}^{\rm SM}=3.044$
\cite{Akita:2020szl,Froustey:2020mcq,Bennett:2020zkv,Drewes:2024wbw}.  We compare it to an extension in which
\begin{equation}
  N_{\rm eff}=3.044+\dNeff .
\end{equation}
A positive $\dNeff$ raises the radiation density throughout BBN.  At fixed
$\wb$, the resulting faster expansion leaves more deuterium unburned, but
also increases the neutron abundance at the onset of nucleosynthesis, and hence $Y_p$.  We allow $\dNeff$ to
take either sign in the analysis in order to map the full likelihood around
its best fit.

\subsection{Very early dark energy}
\label{sec:ede}

\subsubsection{Energy density evolution}

We adopt a phenomenological vEDE density evolution motivated by axion-like
models with a potential $V(\phi)\propto[1-\cos(\phi/f)]^n$
\cite{Kamionkowski:2014zda,Karwal:2016vyq,Poulin:2018dzj,Poulin:2018cxd}.  In these models, the field is initially
frozen and becomes dynamical at the critical redshift $\zc$, after which its
time-averaged equation of state is
\begin{equation}
 w_n=\frac{n-1}{n+1}.
\end{equation}
In the post-$e^\pm$-annihilation regime, where $T\propto a^{-1}$, the vEDE
energy density evolves as
\begin{equation}
 \rho_{\rm vEDE}(T)=
 \frac{2\rho_{\rm vEDE}(T_c)}
 {1+(T_c/T)^{\alpha}},\qquad
 \alpha=3(1+w_n),
 \label{eq:rhoede}
\end{equation}
where $T_c=T_0(1+\zc)$ and $T_0=2.7255\,{\rm K}$~\cite{Fixsen:2009ug}.  We fix $n=3$, for which
$w_n=1/2$ and $\alpha=9/2$, so that the field dilutes faster than radiation
after becoming dynamical. This choice is made by analogy with the
standard EDE parameterization used in CMB analyses of the Hubble tension,
where $n=3$ is a common benchmark that gives a rapidly-diluting
post-transition component \cite{Poulin:2018cxd,Poulin:2023lkg}.
The results are not driven by this specific value: we have checked that using
a much larger $n$, for which $w_n\to1$ and the post-transition evolution
approaches kination, does not qualitatively alter the BBN reconciliation.  We
therefore expect the conclusions to be stable for any post-transition
evolution that dilutes faster than radiation.  They would be affected only if
the additional component diluted more slowly than, or at the same rate as,
radiation, since in that case the expansion-rate enhancement would persist
outside the deuterium-burning window and would no longer be equivalent to the
transient vEDE histories studied here.

The fractional vEDE contribution at the critical redshift $\zc$ is 
\begin{equation}
 \fede(\zc)=
 \frac{\rho_{\rm vEDE}(T_c)}
 {\rho_{\rm std}(T_c)+\rho_{\rm vEDE}(T_c)}.
 \label{eq:fede}
\end{equation}
The total Hubble rate entering the nuclear evolution is
\begin{equation}
 H^2_\mathrm{tot}(T)=\frac{8\pi G}{3}
 \left[\rho_{\rm std}(T)+\rho_{\rm vEDE}(T)\right].
 \label{eq:hubble}
\end{equation}

\subsubsection{vEDE amplitude during BBN}
\label{subsec:pivot}

The conventional parameters $(\fede,\zc)$ are well suited to CMB analyses,
but not to the BBN problem considered here.  BBN is sensitive to the
expansion rate over a finite temperature range, rather than to the vEDE fraction at the critical redshift.  We therefore trade $\fede$ for
\begin{equation}
 \dD\equiv\dH(\TD)
 =\frac{H_{\rm tot}(\TD)}{H_{\rm std}(\TD)}-1 \, ,
 \label{eq:deltaD}
\end{equation}
where $\TD$ is a pivot temperature for which deuterium, but not helium-4, is sensitive to the expansion rate (see below). $\dD$ directly fixes the vEDE density at the pivot:
\begin{equation}
 \rho_{\rm vEDE}(\TD)=
 \left[(1+\dD)^2-1\right]\rho_{\rm std}(\TD).
 \label{eq:rhoD}
\end{equation}
Combining Eqs.~\eqref{eq:rhoede} and \eqref{eq:rhoD} gives
\begin{equation}
 \rho_{\rm vEDE}(T_c)=\frac{1}{2}
 \left[(1+\dD)^2-1\right]\rho_{\rm std}(\TD)
 \left[1+\left(\frac{T_c}{\TD}\right)^\alpha\right].
 \label{eq:mapping}
\end{equation}
For a given $(\dD, z_c)$, the fractional contribution at $z_c$, $\fede(\zc)$, is then a derived parameter,
obtained by combining Eqs.~\eqref{eq:mapping} and \eqref{eq:fede}. 

\begin{figure}
 \centering
    \includegraphics[width=\columnwidth]{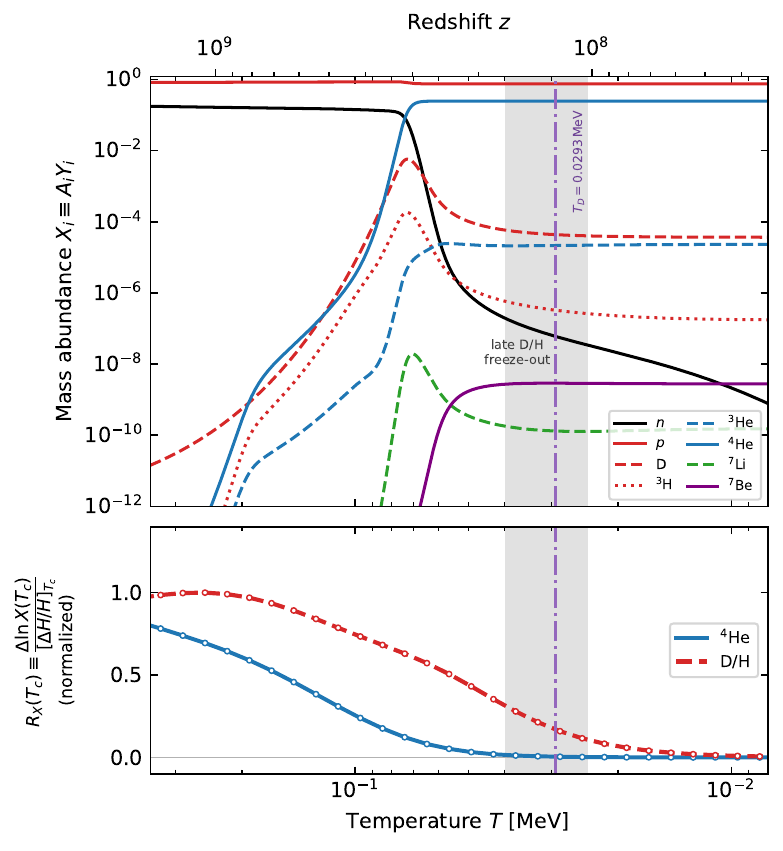}
 \caption{\emph{Top:} Standard evolution of light element abundances obtained with \primat, using a small network of nuclear reactions. $Y_i \equiv n_i/n_B$ is the ratio of the density of $i$ and the total baryon density, and $A_i$ is the mass number, so the plotted mass abundance is $X_i\equiv A_iY_i$; the upper axis gives the corresponding redshift. 
\emph{Bottom:} normalized sensitivity of the final abundances to a localized change in the expansion rate,
 $R_X(T_c)\equiv\Delta\ln X(T_c)/[\Delta H/H]_{T_c}$, where $X$ is a
 final light-element abundance (D/H or $Y_p$), $T_c$ is the temperature at
 which a fixed-amplitude expansion boost is centered, and
 $[\Delta H/H]_{T_c}$ is the realized fractional expansion-rate boost
 actually produced at $T_c$ (open circles: the individual scanned solves,
 each a full nonlinear \primat run differenced against a no-boost baseline). The shaded band highlights the late deuterium-burning region around $\TD$, where changing the expansion rate alters the residual D/H abundance after the bulk of helium-4 synthesis is complete. Increasing $H$ in this window leaves more residual deuterium while acting well after weak freeze-out and the main helium-4 production epoch.}
 \label{fig:evolution}
\end{figure}

The reason for sampling $\dD$ rather than $\fede$ is that BBN is sensitive to
the vEDE density near $\TD$, whereas $\fede$ is defined at the transition
temperature $T_c$.  Whenever $T_c$ lies far from the BBN-sensitive window,
on either side, the same expansion-rate change at $\TD$ maps through
Eq.~\eqref{eq:mapping} onto a peak vEDE fraction that grows rapidly and can
approach unity.\footnote{From Eq.~\eqref{eq:mapping}, at fixed $\dD$ we have $\rho_\mathrm{EDE}(T_c)/\rho_\mathrm{std}(T_c) \propto (\TD/T_c)^4 \left[1 + (T_c/\TD)^\alpha\right]$, using that $\rho_\mathrm{std}(T) \propto T^4$ in the radiation era. Since $\alpha = 9/2$, for $T_c \ll \TD$ or $T_c \gg \TD$ the ratio goes to infinity, such that $f_{\rm vEDE} \to 1$ [see Eq.~\eqref{eq:fede}].}  A flat prior on $\fede$ would therefore spend most of its
volume on this poorly constrained extrapolation, whereas $\dD$ directly
labels the expansion-rate change to which the abundances respond. 
We adopt 
\begin{equation}
 \TD=3.4\times10^8\,{\rm K}=0.0293\,{\rm MeV} \, ,
 \label{eq:TD}
\end{equation}
which corresponds to $z_\mathrm{D}\simeq1.25\times10^8$. This choice should be understood as a parameterization pivot, not a new fundamental scale. 
Its physical motivation is illustrated in Fig.~\ref{fig:evolution}.  The deuterium bottleneck is released earlier, at
$T \simeq0.06$--$0.08\,{\rm MeV}$: deuterium first becomes resistant to
photodissociation, rises rapidly, and opens the path to helium synthesis (note that in this phase, the deuterium density tracks in nuclear statistical equilibrium value).
The residual deuterium abundance is fixed by reactions such as
$\mathrm{D}(p,\gamma)^3{\rm He}$ and the two $\mathrm{D}+\mathrm{D}$ channels, which continue
to destroy deuterium until their rates fall below the Hubble rate at
$T $ of a few $10^{-2}\,{\rm MeV}$.  In a representative standard
BBN calculation, ${\rm D/H}=2.88\times10^{-5}$ at $\TD$, compared with
its asymptotic value $2.44\times10^{-5}$; about 15\% of the deuterium present
at the pivot is therefore still destroyed afterward.

This residual burning is precisely what the pivot is designed to capture:
$\dH(\TD)$ probes the expansion rate while the final D/H is still being
set, but well after neutron--proton weak freeze-out at
$T \sim0.8\,{\rm MeV}$ and the production of most $^4$He (around $T\sim0.07\,{\rm MeV}$, discussed above). It is
therefore considerably more specific to deuterium than an expansion-rate
enhancement maintained throughout BBN. We make this quantitative in the
lower panel of Fig.~\ref{fig:evolution}, which shows the sensitivity
$R_X(T_c)\equiv\Delta\ln X(T_c)/[\Delta H/H]_{T_c}$ of the final abundances to
a localized change in the expansion rate, where $T_c$ is the (fixed-amplitude)
boost's temperature, not the evolving photon temperature of a single history:
each point is a full nonlinear \primat solve with the boost centered at
$T_c$, differenced (in log) against a single no-boost baseline. At $\TD$ the
D/H response is still appreciable, while the $Y_p$ response is already
negligible. This is the desired regime: $\dD$ changes the amount of residual
deuterium destruction without significantly changing helium-4.  Moving the pivot to
$8\times10^8\,{\rm K}\simeq0.069\,{\rm MeV}$ would instead place it at the
bottleneck release, where it overlaps the rapid assembly of $^4$He, whereas
$2.3\times10^8\,{\rm K}\simeq0.020\,{\rm MeV}$ is already close to the final
deuterium plateau.  In practice, we have tested several pivot temperatures
across this window, and the value in Eq.~\eqref{eq:TD} lies safely inside the
deuterium sensitivity window of Fig.~\ref{fig:evolution}. 

\section{Analysis setup}
\label{sec:method}

\subsection{BBN calculation}

We compute $Y_p$ and D/H with the latest \primat code, rewritten in Python (version~0.3.2),\footnote{\primat is available at \url{https://github.com/CyrilPitrou/primat}} which includes the same thermodynamic, weak-rate, and nuclear-network calculation elements described in~\cite{Pitrou:2018cgg,Froustey:2020mcq}. The Bayesian MCMC and profile-likelihood analyses are driven through the \texttt{primat\_tools} interface,\footnote{\texttt{primat\_tools} is available at \url{https://github.com/CyrilPitrou/primat_tools}; it wraps \primat in a \texttt{Cobaya} likelihood and provides the fixed-$\wb$ profile-likelihood driver.} which evaluates the abundance likelihood given below in Eq.~\eqref{eq:chi2bbn} at each sampled or profiled point. The only modification we make to \primat itself is to add the vEDE contribution to the total expansion rate [Eq.~\eqref{eq:hubble}], normalized through the fixed value $\dD\equiv\dH(\TD)=H_{\rm tot}/H_{\rm std}(\TD)-1$ [Eq.~\eqref{eq:deltaD}] at the pivot temperature $\TD$; the conventional fraction $\fede(\zc)$ is then obtained as a derived quantity.  We use the ``small'' network,
which contains eight light nuclides and the twelve key
thermonuclear reactions
$n(p,\gamma)\mathrm{D}$, $\mathrm{D}(p,\gamma)^3{\rm He}$,
$\mathrm{D}(\mathrm{D},n)^3{\rm He}$, $\mathrm{D}(\mathrm{D},p){}^3\mathrm{H}$, $^3\mathrm{H}(p,\gamma)^4{\rm He}$,
${}^3\mathrm{H}(\mathrm{D},n)^4{\rm He}$, $^3\mathrm{H}(^4{\rm He},\gamma)^7{\rm Li}$,
$^3{\rm He}(n,p){}^3\mathrm{H}$, $^3{\rm He}(\mathrm{D},p)^4{\rm He}$, $^3{\rm He}(^4{\rm He},\gamma)^7{\rm Be}$, $^7{\rm Be}(n,p)^7{\rm Li}$, and $^7{\rm Li}(p,^4{\rm He})^4{\rm He}$, in addition to the weak neutron--proton
rates.  This network includes the reactions that control the $Y_p$ and D/H
predictions while allowing \primat to be evaluated directly in the MCMC and
PL analyses. 
The larger network is only relevant for lithium and is considerably
slower, so we use the small network for computational efficiency.
The central weak-rate normalization uses a neutron lifetime
$\tau_n=878.4\,{\rm s}$~\cite{ParticleDataGroup:2024cfk}.
We include the
tabulated \primat Monte Carlo abundance uncertainties from nuclear rates and
the neutron lifetime, added in quadrature to the observational errors.  We then profile (or, in the Bayesian case, marginalize) this effective abundance likelihood over the cosmological model parameters, without explicitly profiling the individual nuclear-rate nuisance parameters.  

\subsection{Datasets and likelihoods}

We model the abundance measurements in Eqs.~\eqref{eq:dhobs} and
\eqref{eq:ypobs} as independent Gaussian likelihoods.  Their contribution to
the total $-2\ln\mathcal{L}_{\rm BBN}$ is
\begin{equation}
 \begin{split}
 -2\ln\mathcal{L}_{\rm BBN}(\bm\theta)&\equiv\chi^2_{\rm BBN}(\bm\theta)\\
 &=\sum_{X}\left[\frac{\big(X-X_{\rm obs}\big)^2}{\sigma_{{\rm tot},X}^2}
 +\ln\!\big(2\pi\,\sigma_{{\rm tot},X}^2\big)\right] ,
 \end{split}
 \label{eq:chi2bbn}
\end{equation}
where the sum runs over $X=Y_p$ and $10^5 \, {\rm D/H}$, with the total variance
$\sigma_{{\rm tot},X}^2(\bm\theta)=\sigma_{{\rm obs},X}^2+\sigma_{{\rm th},X}^2(\bm\theta)$
and observational central values and errors
$(Y_p,\sigma_{\rm obs})=(0.2458,0.0013)$ and
$(10^5 \, {\rm D/H},\sigma_{\rm obs})=(2.527,0.030)$ from
Eqs.~\eqref{eq:ypobs} and~\eqref{eq:dhobs}. Here $\sigma_{{\rm th},X}(\bm\theta)$
is the tabulated \primat Monte Carlo abundance uncertainty from the nuclear
rates and neutron lifetime, added in quadrature to the observational error.
Because $\sigma_{\rm th}$ depends on $\bm\theta$, the Gaussian normalization
term $\ln(2\pi\sigma_{\rm tot}^2)$ cannot be dropped: it penalizes regions of
inflated theoretical error and is required for the likelihood to be properly
normalized, reducing to the usual observational $\chi^2$ (up to an additive
constant) only when $\sigma_{\rm th}$ is independent of $\bm\theta$. The same
$\mathcal{L}_{\rm BBN}$ is minimized in the profile-likelihood analysis and
sampled in the Bayesian analysis. For brevity we write
$\chi^2_{\rm BBN}\equiv-2\ln\mathcal{L}_{\rm BBN}$; the profile-likelihood and
$Q_{\rm DMAP}$ statistics defined below are \emph{differences} of this
quantity, in which the additive normalization constants cancel.
We do not include $^3{\rm He}$ or $^7{\rm Li}$ in the likelihood, the former lacking precision measurements (although recent progress was achieved in~\cite{Cooke:2026juw}), and the latter being related to the well-known ``lithium problem,'' whose status is at present unclear~\cite{ParticleDataGroup:2026}.

When combining BBN with CMB information, we use a one-dimensional Gaussian
likelihood for the baryon density,
\begin{equation}
-2\ln\mathcal{L}_{\rm CMB}(\wb)\equiv\chi^2_{\rm CMB}(\wb)
 =
 \left(\frac{\wb-\mu_{\rm EDE}}{\sigma_{\rm EDE}}\right)^2
 + {\rm const.},
 \label{eq:cmbpriors}
\end{equation}
with
\begin{equation}
\label{eq:constraint_wb}
(\mu_{\rm EDE},\sigma_{\rm EDE})
 = (0.02272,\,0.00014).
\end{equation}
This Gaussian is a fit to the marginalized $\wb$ posterior from an EDE
analysis of the same SPA CMB data set~\cite{Schoneberg:2026eys,Schoneberg:2026vaf}.
We use the same EDE-derived baryon-density likelihood for $\Lambda$CDM,
$\Delta N_{\rm eff}$, and vEDE when combining BBN and CMB information.
This choice is deliberate: the EDE fit provides a representative
high-$\omega_b$ target from early-time Hubble-tension solutions, making the
BBN consistency problem sharper and connecting the possible BBN-era vEDE
episode to the broader sequence of scalar-field epochs discussed above.
This is not meant to assign the EDE CMB likelihood to the other BBN models as
a self-contained cosmology; rather, we ask which BBN expansion histories can
accommodate that target.  In Appendix~\ref{app:lcdmprior} we repeat the
comparison using the lower $\Lambda$CDM CMB baryon-density constraint from the
same SPA data set.

We approximate the joint analysis by simply summing $\chi^2_{\rm CMB+BBN}\equiv \chi^2_{\rm BBN}+\chi^2_{\rm CMB}$ with a {\it shared} $\wb$. Because the vEDE component considered here has diluted away long before recombination, we do not expect it to strongly affect the primary CMB spectra through the usual recombination-era EDE mechanism.   However, Ref.~\cite{Sobotka:2024tat} shows that very early transient components can still leave signatures in the matter transfer function on scales that are inside the horizon while the vEDE density is dynamically important.  Our compressed CMB treatment therefore isolates the baryon-density comparison and does not attempt to model these small-scale matter-power signatures.
Our Gaussian approximation isolates
the comparison of the baryon-density determinations, but does not replace a
joint BBN+CMB analysis.  More precisely, our use of the EDE-derived CMB constraint on
$\wb$ should be viewed as a proxy for a full analysis containing both an
EDE component relevant near recombination and a distinct vEDE component
relevant during BBN. The CMB information fixes the EDE-correlated target
value of $\wb$, while the BBN likelihood constrains vEDE. A joint
EDE+vEDE fit to the complete CMB and abundance likelihoods is beyond the
scope of this work.

To quantify the consistency between the BBN and CMB determinations of
$\wb$, we use the difference in best-fit $\chi^2$ at the maximum a-posteriori (MAP) \cite{Raveri:2018wln}, 
\begin{align}
 \Delta\chi^2_{\rm DMAP}
 &=\chi^2_{\rm MAP,BBN+CMB}
 -\chi^2_{\rm MAP,BBN}
 -\chi^2_{\rm MAP,CMB},\nonumber\\
 Q_{\rm DMAP}&\equiv\sqrt{\Delta\chi^2_{\rm DMAP}}\, .
 \label{eq:qdmap}
\end{align}
For the compressed Gaussian CMB likelihood used here,
$\chi^2_{\rm MAP,CMB}=0$ by construction. For one effective degree of
freedom, $Q_{\rm DMAP}$ is the corresponding tension in units of standard
deviations. Since the target is the consequence of an early-time
Hubble-tension solution---namely the elevated CMB baryon density---we evaluate
both $Q_{\rm DMAP}$ and $\Delta\mathrm{AIC}$ for all three BBN models against
the common EDE CMB constraint from Eq.~\eqref{eq:constraint_wb}.
Note that $Q_{\rm DMAP}$ is a dataset-compatibility (goodness-of-fit) metric, not a model-selection criterion: it carries no explicit penalty for the number of extra parameters. We perform model selection by computing the Akaike information
criterion, $\mathrm{AIC}=\chi^2_{\rm min}+2k$, where $k$ is the number of parameters in a model, evaluated on the combined
BBN$+$CMB fit. We report the difference between $\Lambda$CDM and the extended models, $\Delta {\rm AIC}\equiv \Delta \chi^2+2\Delta k$, where $\Delta k=1$ for $\Delta N_{\rm eff}$ and $\Delta k=2$ for vEDE.

\subsection{Statistical analysis}

To obtain constraints on $\wb$, $\Delta N_{\rm eff}$ and $\{\dD,z_c\}$, we perform a Bayesian MCMC analysis with \texttt{Cobaya}~\cite{Torrado:2020dgo}, running four chains in parallel.  We consider the chains to be
converged when they fulfill the Gelman--Rubin criterion $R-1<0.01$.  For vEDE,
we adopt uniform priors
\begin{equation}
\begin{aligned}
 0.0205&<\wb<0.0240 \, ,\qquad 0<\dD<0.20 \, ,\\
 7&<\log_{10}\zc<10 \, ,
\end{aligned}
\end{equation}
while for the radiation extension we use
\begin{equation}
 0.0205<\wb<0.0240 \, ,\qquad -0.5<\dNeff<0.6 \, .
\end{equation}
For each extension we perform dedicated BBN-only and BBN+CMB analyses. 

We complement the Bayesian analysis with a frequentist PL analysis of the
BBN data.  At each of $\sim 20$ fixed $\wb$ values in the range
defined above, we minimize Eq.~\eqref{eq:chi2bbn} over
$\{\dD,\log_{10}\zc\}$ for vEDE and over $\{\dNeff\}$ for the radiation
extension. 
MCMC points provide the initial positions for \textsc{Minuit}, a numerical minimization and error-analysis package~\cite{James:1975dr}, followed by a bounded numerical minimization. This procedure determines the best fit at each fixed $\wb$.

For each model, we define the BBN-only PL as
\begin{equation}
 \Delta\chi^2_{\rm BBN,PL}(\wb)\equiv
 \chi^2_{\rm BBN}\!\left(\wb,\widehat{\bm\theta}(\wb)\right)
 -\chi^2_{\rm BBN,min},
 \label{eq:compat}
\end{equation}
where $\widehat{\bm\theta}(\wb)$ denotes the model-specific parameters  that
minimize $\chi^2_{\rm BBN}$ at fixed $\wb$ (for $\Lambda$CDM the likelihood is simply evaluated at a given $\wb$, while we vary $\{\dNeff\}$ for the radiation extension and $\{\dD,\log_{10}z_c\}$ for vEDE).

\section{Results}
\label{sec:results}
Results of the different analyses are presented in Table~\ref{tab:posterior}. We provide posterior means, best fits and credible intervals at 68\% for cosmological parameters and BBN yields, as well as the tension metric $Q_{\rm DMAP}$ and the model-selection statistic $\Delta{\rm AIC}$.

\renewcommand{\arraystretch}{1.05}
\begin{table*}
    \resizebox{1\textwidth}{!}{%
\begin{ruledtabular}
\begin{tabular}{lcc}
Parameter & BBN & BBN+CMB $\wb$\\
\hline
\multicolumn{3}{c}{$\bm{\Lambda}$\textbf{CDM}}\\
$100\,\wb$ &
$2.1925^{+0.0212}_{-0.0211}\ (2.1882)$ &
$2.2479^{+0.0118}_{-0.0113}\ (2.2488)$\\
$Y_p$ &
$0.246781^{+0.000093}_{-0.000093}\ (0.246759)$ &
$0.247022^{+0.000051}_{-0.000049}\ (0.247024)$\\
$10^5 \,{\rm D/H}$ &
$2.5272^{+0.0397}_{-0.0399}\ (2.5346)$ &
$2.4256^{+0.0201}_{-0.0210}\ (2.4237)$\\
$Q_{\rm DMAP}$ & -- & $3.1\sigma$\\
$\Delta{\rm AIC}$ & -- & $0$\\
\hline
\multicolumn{3}{c}{$\bm{\Delta N_{\rm eff}}$}\\
$100\,\wb$ &
$2.1817^{+0.0255}_{-0.0265}\ (2.1727)$ &
$2.2518^{+0.0118}_{-0.0118}\ (2.2523)$\\
$\Delta N_{\rm eff}$ &
$-0.064^{+0.094}_{-0.093}\ (-0.083)$ &
$0.076^{+0.076}_{-0.079}\ (0.078)$\\
$Y_p$ &
$0.245863^{+0.001323}_{-0.001325}\ (0.245570)$ &
$0.248048^{+0.001034}_{-0.001059}\ (0.248071)$\\
$10^5 \, {\rm D/H}$ &
$2.5255^{+0.0412}_{-0.0404}\ (2.5354)$ &
$2.4434^{+0.0285}_{-0.0273}\ (2.4428)$\\
$Q_{\rm DMAP}$ & -- & $3.1\sigma$\\
$\Delta{\rm AIC}$ & -- & $\simeq+1.2$\\
\hline
\multicolumn{3}{c}{\textbf{vEDE}}\\
$100\,\wb$ &
$2.2650^{+0.0535}_{-0.0531}\ (2.1882)$ &
$2.2708^{+0.0141}_{-0.0137}\ (2.2713)$\\
$\dD=\Delta H/H(\TD)$ &
$0.082^{+0.071}_{-0.065}\ (0.000)$ &
$0.087^{+0.036}_{-0.037}\ (0.079)$\\
$Y_p$ &
$0.247208^{+0.000257}_{-0.000302}\ (0.246758)$ &
$0.247218^{+0.000063}_{-0.000111}\ (0.247156)$\\
$10^5 \, {\rm D/H}$ &
$2.5308^{+0.0388}_{-0.0395}\ (2.5346)$ &
$2.5292^{+0.0374}_{-0.0391}\ (2.5272)$\\
$Q_{\rm DMAP}$ & -- & $0.7\sigma$\\
$\Delta{\rm AIC}$ & -- & $\simeq-5.3$\\
\end{tabular}
\end{ruledtabular}
}
\caption{\small Posterior means and equal-tailed 68\% Bayesian
credible intervals, with best-fit values given in parentheses. The
BBN+CMB columns use the CMB-derived value of the baryon-density in the EDE model as a Gaussian likelihood for
all three models. We also report the tension metric $Q_{\rm DMAP}$ defined in
Eq.~\eqref{eq:qdmap} and $\Delta{\rm AIC}$, referenced to $\Lambda$CDM.
\label{tab:posterior}}
\end{table*}

\subsection{BBN profile likelihoods}

\begin{figure}[t]
 \centering
 \includegraphics[width=\columnwidth]
 {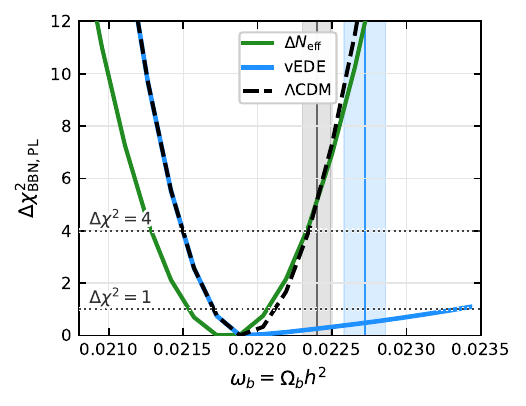}
 \caption{BBN-only PL of $\wb$, as defined in Eq.~\eqref{eq:compat}, within $\Lambda$CDM (dashed black), $\dNeff$
 (solid green), and vEDE (solid blue), normalized to the best fit of each model.  The
 vertical band shows the common EDE CMB constraint from Eq.~\eqref{eq:constraint_wb}.
 It is shown only as a reference; the CMB likelihood is not included in the PL.}
 \label{fig:profiles}
\end{figure}

We show the BBN-only PL of $\wb$ in Fig.~\ref{fig:profiles}.  The BBN-only
profile best fits and 68\% confidence intervals, obtained from a smooth fit/interpolation of
the tabulated fixed-$\wb$ profile points around the minimum and defined by
$\Delta\chi^2_{\rm BBN,PL}\leq1$, are
\begin{equation}
\Lambda{\rm CDM}:\qquad
\wb=0.021882^{+0.000247}_{-0.000186}\quad(68\%~{\rm CL}) \, .
 \label{eq:standardprofile}
\end{equation}
For the two extensions we find
\begin{align}
 \dNeff:\qquad
 &\wb=0.021727^{+0.000345}_{-0.000197}\quad(68\%~{\rm CL}),
 \nonumber\\
 \mathrm{vEDE}:\qquad
 &\wb=0.021882^{+0.001464}_{-0.000186}\quad(68\%~{\rm CL}).
 \label{eq:extensionprofiles}
\end{align}

Allowing $\dNeff$ broadens slightly the BBN confidence interval towards larger
$\wb$.  However, the PL rises rapidly across the CMB-preferred region. 
The
vEDE result is qualitatively different: its PL remains nearly flat across the
EDE CMB constraint, showing that the vEDE enables compatible high-$\omega_b$ and high-D/H values.

Within $\Lambda$CDM, the joint BBN+CMB best fit is
$\wb=0.02249$ and $\Delta\chi^2_{\rm DMAP}\simeq9.8$, corresponding to
$Q_{\rm DMAP}=3.1\sigma$.  Allowing $\dNeff$ does not improve the
consistency, yielding $Q_{\rm DMAP}=3.1\sigma$.  By contrast, vEDE gives
$Q_{\rm DMAP}=0.7\sigma$, showing that the BBN and CMB determinations of
$\wb$ are statistically consistent.
We find
$\Delta\mathrm{AIC}\simeq+1.2$ for the one-parameter $\dNeff$ extension and
$\Delta\mathrm{AIC}\simeq-5.3$ for the two-parameter vEDE model. The radiation
extension does not reconcile the two $\wb$ determinations, remaining in $\simeq3\sigma$
tension, whereas vEDE is strongly preferred over $\Lambda$CDM, confirming that its
resolution of the tension is not simply an artefact of the additional
parameters.

\begin{figure}[t!]
 \centering
 \includegraphics[width=\columnwidth]
 {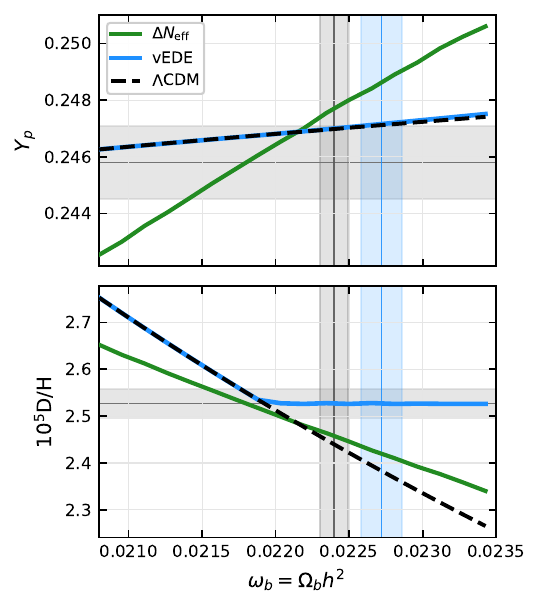}
 \caption{\primat abundance predictions at the best fit of each fixed-$\wb$
 PL in Fig.~\ref{fig:profiles}, for $\Lambda$CDM (dashed black), $\dNeff$ (solid green),
 and vEDE (solid blue).  The horizontal grey bands show the 68\% abundance
 constraints~\eqref{eq:dhobs} and \eqref{eq:ypobs}.  The vertical grey band
 shows the $\Lambda$CDM CMB-SPA determination of $\wb$, while the vertical
 blue band shows the common EDE CMB determination of $\wb$~\eqref{eq:constraint_wb}.}
 \label{fig:abundances}
\end{figure}

These values correspond to the baseline deuterium measurement in
Eq.~\eqref{eq:dhobs}.  Since the \primat prediction at the CMB-preferred
baryon density lies below the observed D/H, adopting the more recent central
value $10^5\,{\rm D/H}=2.508$~\cite{ParticleDataGroup:2026}
weakens the discrepancy: re-evaluating the stored profiles gives
approximately $Q_{\rm DMAP}=2.7\sigma$, $2.8\sigma$, and $0.7\sigma$ for
$\Lambda$CDM, $\dNeff$, and vEDE, respectively, leaving the vEDE result
essentially unchanged.  The dedicated Bayesian analyses with this deuterium
determination are reported in Appendix~\ref{app:newdh}.

\subsection{Impact on the light-element abundances}

To understand these differences, we show in Fig.~\ref{fig:abundances} the
abundances at the best fit of each fixed-$\wb$ PL. At the central value
of the common EDE CMB baryon-density constraint, the baseline calculation
predicts
$Y_p\simeq0.24712$ and $10^5 \, {\rm D/H}\simeq2.383$.  Instead, at the $\dNeff$ BBN+CMB profile best fit, the
abundances are
\begin{equation}
 Y_p=0.24807^{+0.00033}_{-0.00036},\qquad
 10^5 \, {\rm D/H}=2.443^{+0.015}_{-0.015}.
\end{equation}
Relative to $\Lambda$CDM, the positive $\dNeff$ shift does move
D/H toward the observed value
$2.527$. This improvement is too small to compensate for the simultaneous
increase in $Y_p$ induced by the enhanced expansion rate during the neutron--proton weak freeze-out, so the combined abundance fit is worse. 

vEDE instead localizes the expansion-rate enhancement around the
deuterium-burning epoch. The prediction at the vEDE BBN+CMB profile best fit for the
abundances is
\begin{equation}
 Y_p=0.24716^{+0.00007}_{-0.00007},\qquad
 10^5 \, {\rm D/H}=2.527^{+0.001}_{-0.001}.
\end{equation}

The faster expansion during deuterium freeze-out compensates for the more
efficient deuterium burning at larger $\wb$.  Since the vEDE contribution is
transient, the corresponding change in $Y_p$ remains small.

\begin{figure*}[!t]
    \centering
        \centering
       \includegraphics[width=\columnwidth]
 {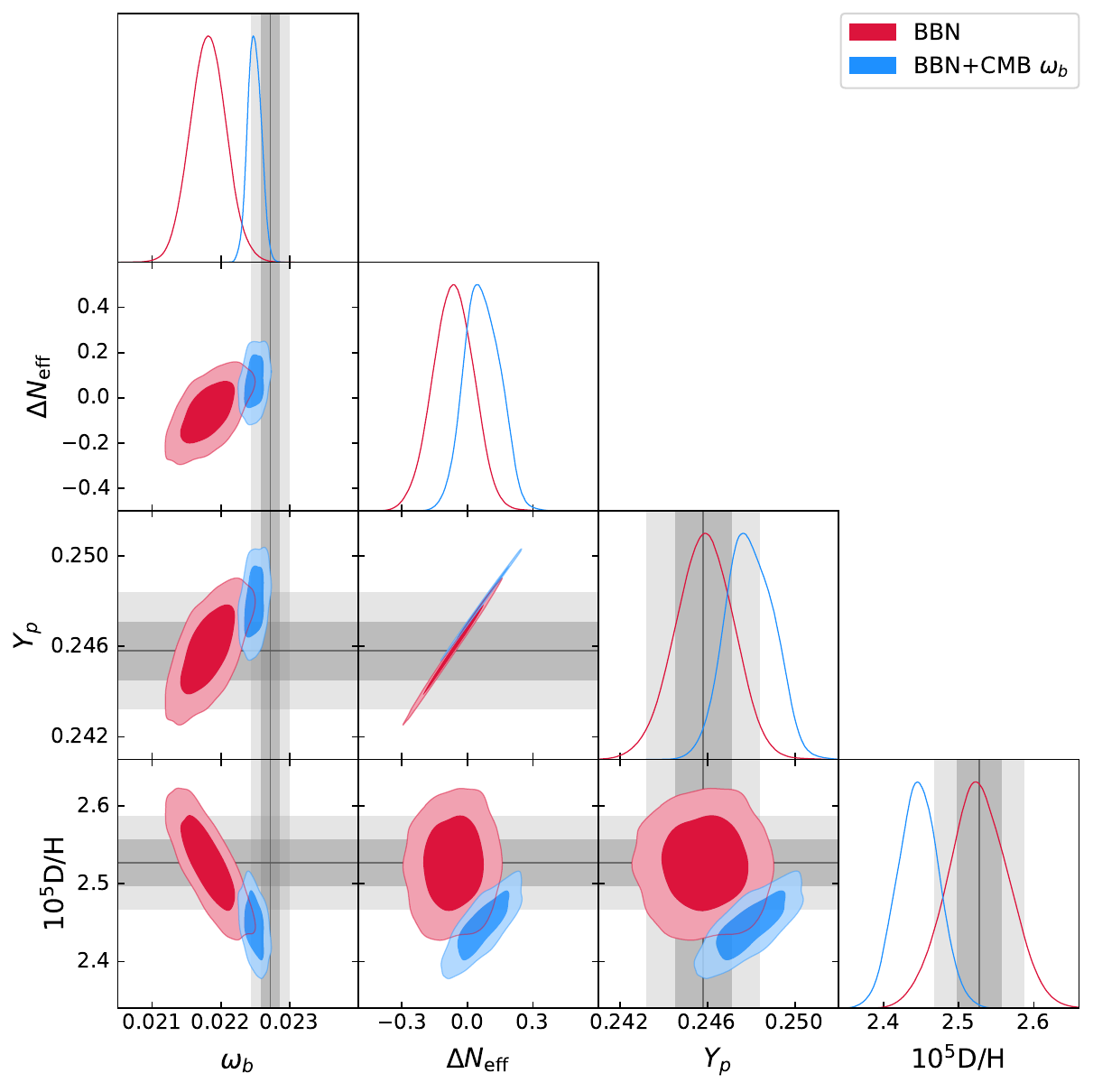}
    \hfill
        \centering
         \includegraphics[width=\columnwidth]
 {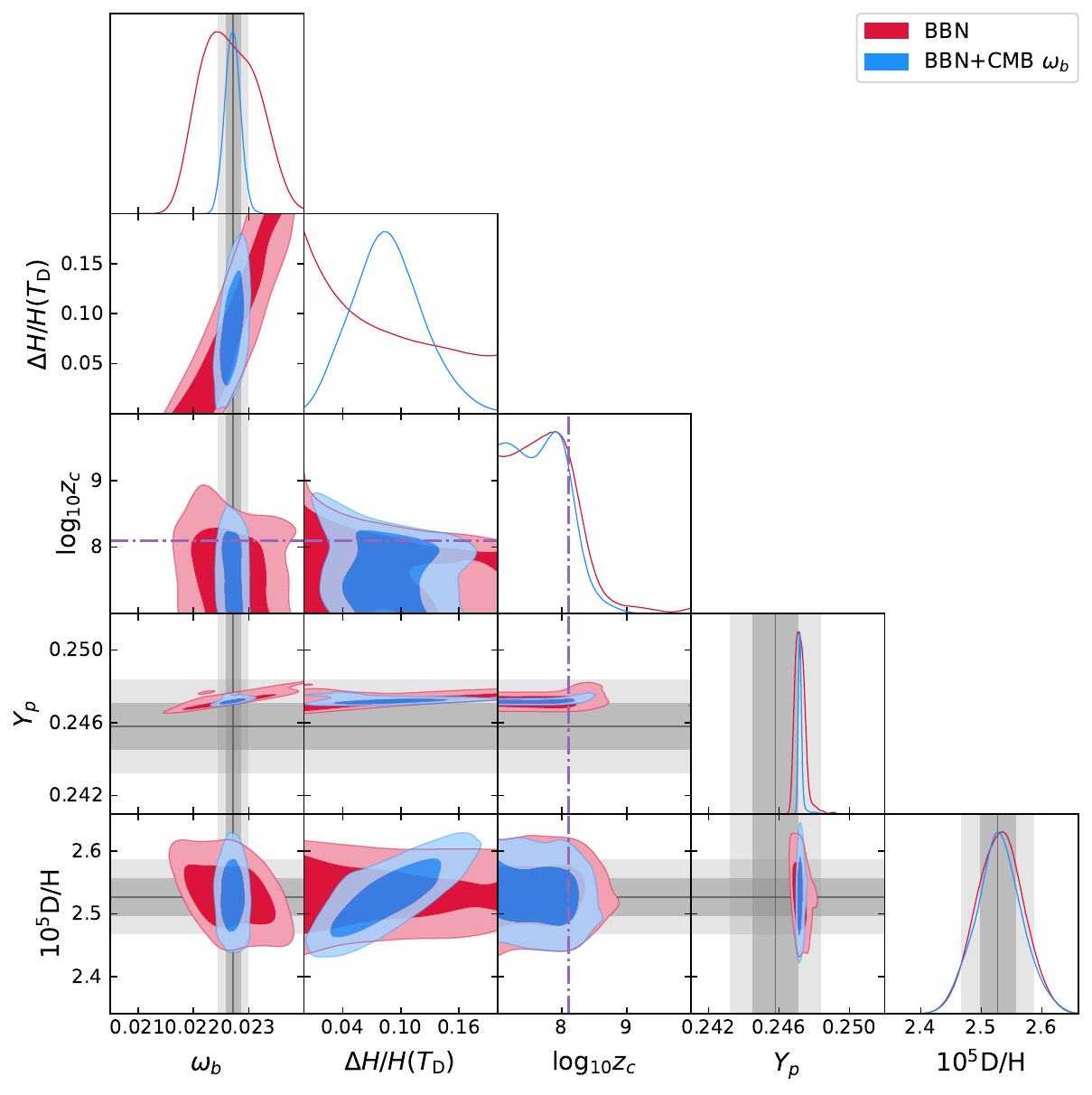}
    \caption{\emph{Left:} Marginalized posterior distributions within the $\dNeff$ model from BBN (red) and
 BBN combined with the EDE CMB $\wb$ constraint (blue). The dark and light contours contain 68\% and 95\% of the posterior probability,
 respectively.  Grey bands show the $1\sigma$ abundance measurements and the
 EDE CMB constraint on $\wb$. \emph{Right}: Same as the left panel but for the vEDE model. The dark
 and light contours contain 68\% and 95\% of the posterior probability,
 respectively.  Grey bands show the $1\sigma$ abundance measurements and the
 EDE CMB constraint on $\wb$. The dash-dotted purple line on the $\zc$ panels corresponds to $z_\mathrm{D}$ [see after Eq.~\eqref{eq:TD}]. }
    \label{fig:combined}
\end{figure*}

\subsection{Bayesian posteriors}

The corresponding marginalized distributions within the $\dNeff$ extension
are shown in the left panel of Fig.~\ref{fig:combined}.  Combining BBN with the
EDE CMB determination of $\wb$ shifts $\dNeff$ to positive values,
with
\begin{equation}
\dNeff=0.076^{+0.076}_{-0.079}
\end{equation}
and predicted abundances
\begin{equation}
Y_p=0.24805^{+0.00103}_{-0.00106},\qquad
 10^5 \, {\rm D/H}=2.443^{+0.028}_{-0.027}.
\end{equation}
The posterior therefore compromises between the CMB baryon density and the
two abundance measurements, consistent with the rapid degradation of the
BBN-only PL across the CMB-preferred region.

The marginalized vEDE posterior distributions are shown in the right panel of
Fig.~\ref{fig:combined}.
The BBN-only vEDE posterior is broad, since a range
of expansion histories produces the required change in deuterium. 
Combining BBN with the EDE CMB determination of $\wb$, we find the posterior
means and equal-tailed 68\% credible intervals
\begin{equation}
\begin{aligned}
 \wb&=0.022708^{+0.000141}_{-0.000137}, \\
 \dD&=0.087^{+0.036}_{-0.037}\,, \label{eq:edepost}
\end{aligned}
\end{equation}
while $\log_{10}\zc < 8.3$ (95\% C.L.).
The corresponding predicted abundances are
$Y_p=0.24722^{+0.00006}_{-0.00011}$ and
$10^5 \, {\rm D/H}=2.529^{+0.037}_{-0.039}$. This latter value is now in perfect agreement with~\eqref{eq:dhobs}.

The CMB constraint selects the positive
correlation between $\wb$ and $\dD$: a larger baryon density requires a
larger expansion-rate enhancement to preserve deuterium.  By contrast, the
posterior remains broad in $\zc$.  Through Eq.~\eqref{eq:mapping}, this broad
range of critical redshifts corresponds to very different values of
$\fede(\zc)$ but similar expansion histories during deuterium freeze-out. In Appendix~\ref{app:fvEDE}, we further discuss the posterior of the derived quantity $\fede(z_c)$, which suffers from prior-volume effects, contrary to the physically-sound sampled parameter $\dD$.

\section{Summary and conclusions}
\label{sec:conclusion}

In this paper, we have investigated whether a non-standard expansion history during BBN can reconcile the measured light-element abundances with their predicted values at the high baryon density selected by EDE CMB fits.  We use EDE as a representative example of the baryon-density shift induced by early-time resolutions of the Hubble tension \cite{Giovanetti:2026aku}, and ask what additional BBN-era physics would then be required.   We have compared $\Lambda$CDM, a $\dNeff$ extension, and vEDE (a transient very early dark energy component) using a common \primat abundance likelihood.  We have performed both a
Bayesian MCMC analysis and a frequentist PL analysis of the BBN data.

Within $\Lambda$CDM, the BBN-only PL favors $\wb=0.02188$, with
$0.02170<\wb<0.02213$ at 68\% CL.  Its consistency with the CMB
determination is $Q_{\rm DMAP}=3.1\sigma$.  Allowing constant extra
radiation does not improve the agreement: the faster expansion raises $Y_p$
too efficiently while D/H remains below its measured value, giving
$Q_{\rm DMAP}=3.1\sigma$.

The result changes qualitatively within vEDE.  The BBN-only PL remains nearly
flat across the CMB-preferred range of $\wb$, reducing the discrepancy to
$Q_{\rm DMAP}=0.7\sigma$.  The reason is the timing of the expansion-rate
enhancement.  After the so-called ``deuterium bottleneck,'' a larger expansion
rate around a few $10^{-2}\,{\rm MeV}$ shortens the remaining
deuterium-burning phase.  vEDE can therefore preserve D/H at larger $\wb$
while producing only a mild change in helium-4.  By contrast, $\dNeff$ speeds up the expansion throughout BBN: the increased $H$ leads to a weak freeze-out at higher temperature, and leaves less time for neutron decay, thus resulting in a higher $Y_p$, while the shortened deuterium-burning phase simultaneously raises D/H. The virtue of a transient vEDE is that it accelerates the expansion \emph{only} at low temperatures, around deuterium burning, and not throughout: it likewise shortens the deuterium-burning phase (raising D/H) but leaves weak freeze-out and the bulk of $^4$He synthesis essentially untouched, so $Y_p$ barely moves. 

\begin{figure}
 \centering
 \includegraphics[width=\columnwidth]
 {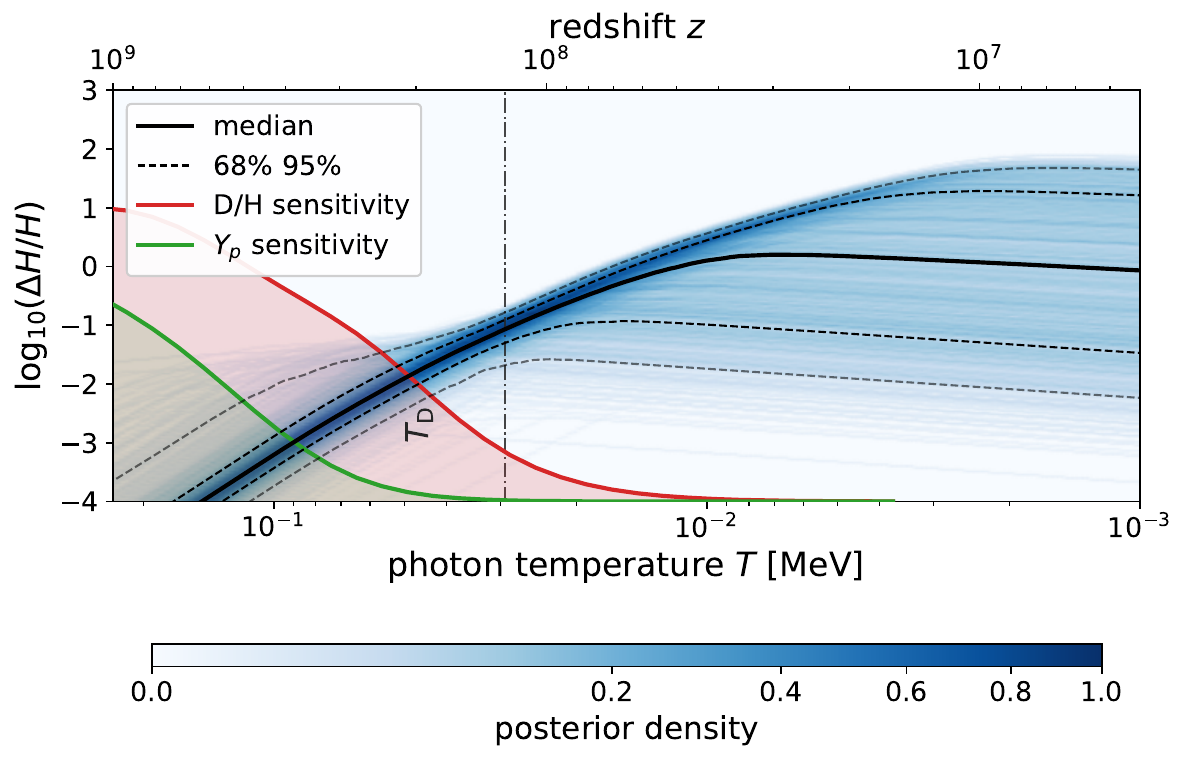}
 \caption{Posterior distribution of the fractional expansion-rate enhancement
 $\dH$ as a function of photon temperature (bottom axis) and redshift (top
 axis), for the combined BBN+CMB analysis with
 $10^5\,{\rm D/H}=2.527\pm0.030$. The blue shading is the marginalized
 posterior density of $\log_{10}\dH$ at each epoch (normalized to unit peak in
 each column); the solid black line is the median and the dashed lines the
 $68\%$ and $95\%$ credible bands. The dotted line marks $\dH=1\%$ and the
 dot-dashed line the pivot $\TD$. Overlaid in red and green are the deuterium
 and helium-4 sensitivity windows $R_X$ of Fig.~\ref{fig:evolution}, arbitrary
 normalized for visual purpose. Data are constraining only around $\TD$, where D/H and
 $Y_p$ actually respond.  The large low-redshift amplitude is the assumed
 transient shape extrapolated into a region to which BBN is insensitive.}
 \label{fig:dHz}
\end{figure}

Importantly, BBN does not directly constrain the vEDE fraction at the critical redshift, $\fede(\zc)$.  At fixed $\dH(\TD)$, the value of $\fede(\zc)$ can approach
unity for critical redshifts on either side of the BBN-sensitive window.
Sampling a flat prior on $\fede$ would therefore introduce large
prior-volume effects unrelated to the expansion rate probed by the
abundances.  We instead use $\dH(\TD)$ at
$\TD=3.4\times10^8\,{\rm K}$, which lies on the late deuterium-burning tail.
Combining BBN with the EDE CMB baryon-density constraint gives
\begin{equation}
 \dH(\TD)=0.087^{+0.036}_{-0.037},\qquad
\end{equation}
with 68\% Bayesian credible intervals; relative to $\Lambda$CDM evaluated at the same EDE CMB baryon density, vEDE is preferred with $\Delta {\rm AIC} \simeq -5.3$, i.e., there is a hint of non-standard expansion at BBN.
This is further illustrated in Fig.~\ref{fig:dHz}, where we show the full posterior of
the fractional expansion-rate enhancement $\dH$. 
Overlaid are the
deuterium and helium-4 sensitivity windows of Fig.~\ref{fig:evolution}
(normalized to unit peak): the posterior enhancement is pinned down by the data around $\TD$, where D/H and
$Y_p$ actually respond.
For comparison, a constant radiation increase modeled as $\Delta N_{\rm eff}$ is not favored over $\Lambda$CDM with $\Delta {\rm AIC} \simeq +1.2$.

Our analysis makes several simplifying assumptions.  We propagate the
\primat nuclear-rate and neutron-lifetime uncertainties through the
tabulated Monte Carlo abundance errors, rather than sampling each
reaction-rate nuisance parameter explicitly.  
We also compress the
CMB information into the EDE-derived Gaussian constraint on $\wb$.  
A full
joint CMB+BBN analysis may reveal correlations with the remaining
cosmological parameters, due to effects of the vEDE components on the small scale matter power spectrum \cite{Sobotka:2024tat} and the very slightly larger $Y_p$.
Our results show that the BBN--CMB baryon-density tension is sensitive not only to the amount of additional energy, but also to when it contributes to the expansion rate.  A transient contribution near deuterium freeze-out provides a concrete way of reconciling the large baryon density predicted by early-time resolutions of the Hubble tension with the primordial light-element abundances. Such scenarios necessarily involve several extra fields compared to the baseline $\Lambda$CDM model. However, one can instead interpret this as different cosmological epochs opening distinct temporal windows on a multi-field dark sector like the {\it axiverse} scenario~\cite{Arvanitaki:2009fg,Kamionkowski:2014zda,Karwal:2016vyq}: the CMB can probe an EDE component active near recombination, whereas primordial abundances probe a much earlier vEDE component around deuterium freeze-out.  Inflation~\cite{Marsh:2013taa} and late-time dark energy~\cite{Kamionkowski:2014zda} may also be caused by such axiverse-type fields. In this sense, vEDE need not be viewed merely as an additional isolated field, but as a new BBN window on an axiverse-like spectrum of fields. The present work does not directly fit or resolve the Hubble tension; it uses the EDE-motivated CMB baryon density as the target of the consistency test. The natural extension is a joint multi-field analysis spanning BBN, the CMB, and late-time distance-ladder data.

\begin{acknowledgments}

We thank Adam Riess for suggesting the idea that motivated this work and for valuable feedback on the manuscript.
We thank Bryce Cyr and Lloyd Knox for useful discussions. 
J.F. acknowledges support from the Severo Ochoa Excellence Grant CEX2023-001292-S funded by MICIU/AEI/10.13039/501100011033.
V.P. is supported by funding from the European Research Council (ERC) under the European Union's HORIZON-ERC-2022 (grant agreement no. 101076865). V.P.\ acknowledges the European Union's Horizon Europe research and innovation programme under the Marie Sk\l odowska-Curie Staff Exchange grant agreement no.\ 101086085 -- ASYMMETRY. The authors acknowledge the use of Anthropic's Claude AI and OpenAI Codex as research-assistance tools for code development, plotting, data analysis, and text editing. The scientific ideas, analysis choices, interpretation of the results, manuscript writing are the responsibility of the authors. The code, data, and notebooks needed to reproduce the figures will be made public upon acceptance of the paper for publication at \href{https://zenodo.org/records/21494819}{https://zenodo.org/records/21494819}. We also acknowledge the use of the Python packages \texttt{NumPy}, \texttt{SciPy}, \texttt{Matplotlib}, \texttt{GetDist}, and \texttt{Cobaya}.
\end{acknowledgments}

\appendix

\section{Results with the updated deuterium abundance}
\label{app:newdh}
\label{app:dh2508}

The baseline analysis presented in the main text uses the Cooke \emph{et al.} determination in Eq.~\eqref{eq:dhobs}, both because it permits a direct comparison with
Ref.~\cite{Giovanetti:2026aku} and because it has been widely used in recent
BBN analyses.  Here we repeat the Bayesian analysis using the more recent
recommended abundance~\cite{Kislitsyn:2024jvk,Schoneberg:2024ifp,ParticleDataGroup:2026}
\begin{equation}
 10^5\,{\rm D/H}=2.508\pm0.030 \, .
 \label{eq:dhobs2508}
\end{equation}
We leave the helium-4 likelihood,
the \primat theoretical uncertainties, all parameter priors, and the
EDE-derived CMB baryon-density constraint unchanged.

\begin{figure*}[t!]
 \centering
 \includegraphics[width=\textwidth]
{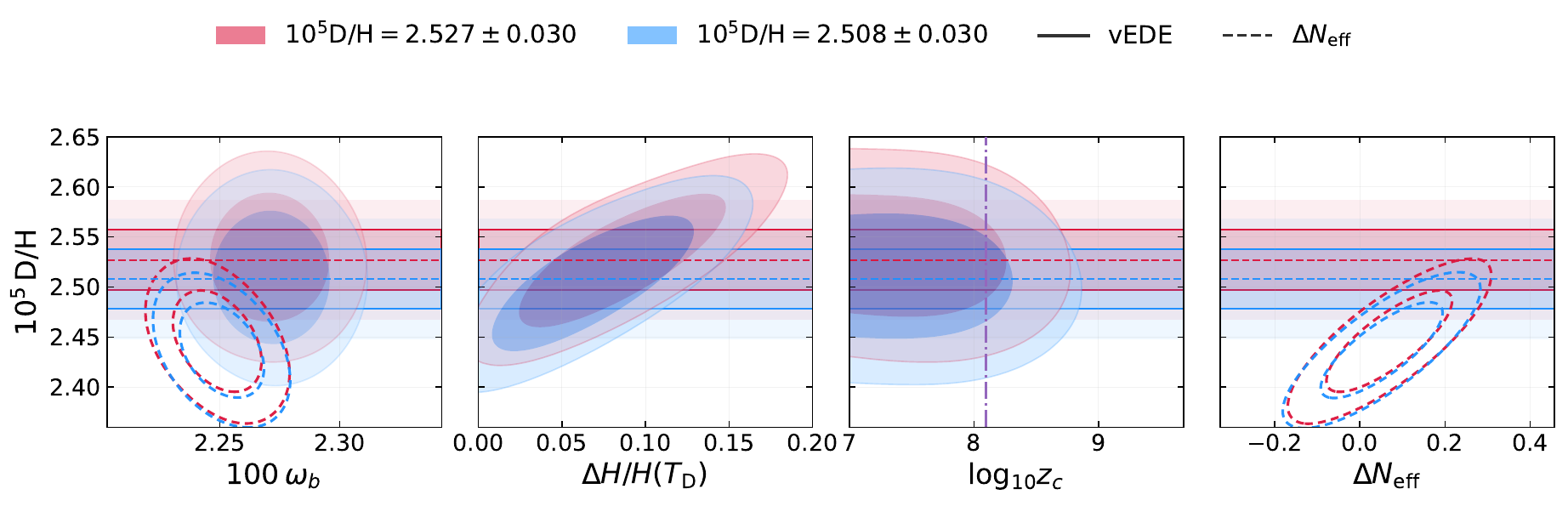}
 \caption{Joint BBN+CMB posterior distributions of D/H against the parameters
 that control the two extensions, comparing the baseline deuterium
 determination $10^5 \, {\rm D/H}=2.527\pm0.030$ (red) with the updated value
 $2.508\pm0.030$ (blue).  Contours enclose 68\% and 95\% of the posterior.
 In the common $100\wb$ panel, filled contours denote vEDE and dashed
 contours denote the $\Delta N_{\rm eff}$ extension; the remaining panels
 show the corresponding model-specific parameters.  Horizontal dashed lines
 and lightly shaded bands indicate the central values and $1\sigma$ ranges of
 the two deuterium determinations.}
 \label{fig:dhpriorcomparison}
\end{figure*}

The dedicated-chain results are reported in
Table~\ref{tab:posterior2508}.  The lower observed deuterium abundance shifts
the BBN-only baryon-density posterior upward, as expected from
${\rm D/H}\propto\wb^{-1.65}$.  For the radiation extension, the BBN+CMB
posterior becomes
\begin{equation}
\wb=0.022543^{+0.000128}_{-0.000133},\qquad
 \dNeff=0.056^{+0.075}_{-0.079}.
\end{equation}
For vEDE we instead find
\begin{align}
 \wb&=0.022716^{+0.000137}_{-0.000137},\nonumber\\
 \dD&=0.073^{+0.035}_{-0.036},\nonumber\\
 \log_{10}\zc&<8.4.
\end{align}
Relative to the baseline result in Eq.~\eqref{eq:edepost}, the preferred
local expansion-rate enhancement is therefore slightly smaller, while the
transition-redshift posterior remains bounded from above.
The effect of the deuterium determination on the joint BBN+CMB posteriors is
shown directly in Fig.~\ref{fig:dhpriorcomparison}.

\begin{table*}[t]
\begin{ruledtabular}
\begin{tabular}{lcc}
Parameter & BBN & BBN+CMB $\wb$\\
\hline
\multicolumn{3}{c}{$\bm{\Lambda}$\textbf{CDM}}\\
$100\,\wb$ &
$2.2017^{+0.0221}_{-0.0217}\ (2.2017)$ &
$2.2519^{+0.0113}_{-0.0123}\ (2.2517)$\\
$Y_p$ &
$0.246821^{+0.000097}_{-0.000095}\ (0.246821)$ &
$0.247039^{+0.000049}_{-0.000053}\ (0.247039)$\\
$10^5 \, {\rm D/H}$ &
$2.5100^{+0.0405}_{-0.0413}\ (2.5094)$ &
$2.4185^{+0.0218}_{-0.0200}\ (2.4187)$\\
$Q_{\rm DMAP}$ & -- & $\simeq2.7\sigma$\\
$\Delta{\rm AIC}$ & -- & $0$\\

\hline
\multicolumn{3}{c}{$\bm{\Delta N_{\rm eff}}$}\\
$100\,\wb$ &
$2.1906^{+0.0262}_{-0.0260}\ (2.1902)$ &
$2.2543^{+0.0128}_{-0.0133}\ (2.2544)$\\
$\Delta N_{\rm eff}$ &
$-0.070^{+0.093}_{-0.091}\ (-0.071)$ &
$0.056^{+0.075}_{-0.079}\ (0.059)$\\
$Y_p$ &
$0.245824^{+0.001317}_{-0.001309}\ (0.245814)$ &
$0.247787^{+0.001018}_{-0.001069}\ (0.247841)$\\
$10^5 \, {\rm D/H}$ &
$2.5068^{+0.0385}_{-0.0416}\ (2.5068)$ &
$2.4324^{+0.0281}_{-0.0285}\ (2.4333)$\\
$Q_{\rm DMAP}$ & -- & $\simeq2.8\sigma$\\
$\Delta{\rm AIC}$ & -- & $\simeq+1.8$\\
\hline
\multicolumn{3}{c}{\textbf{vEDE}}\\
$100\,\wb$ &
$2.2751^{+0.0528}_{-0.0519}\ (2.2005)$ &
$2.2716^{+0.0137}_{-0.0137}\ (2.2718)$\\
$\dD=\Delta H/H(\TD)$ &
$0.081^{+0.070}_{-0.064}\ (0.000)$ &
$0.073^{+0.035}_{-0.036}\ (0.065)$\\
$Y_p$ &
$0.247252^{+0.000251}_{-0.000289}\ (0.246817)$ &
$0.247224^{+0.000061}_{-0.000120}\ (0.247156)$\\
$10^5 \, {\rm D/H}$ &
$2.5104^{+0.0395}_{-0.0397}\ (2.5128)$ &
$2.5094^{+0.0384}_{-0.0379}\ (2.5063)$\\
$Q_{\rm DMAP}$ & -- & $\simeq0.7\sigma$\\
$\Delta{\rm AIC}$ & -- & $\simeq-3.0$\\
\end{tabular}
\end{ruledtabular}
\caption{\label{tab:posterior2508}Same as in Table~\ref{tab:posterior}, but using
the updated deuterium abundance in Eq.~\eqref{eq:dhobs2508}. The
BBN+CMB columns use the CMB-derived value of the baryon-density in the EDE model as a Gaussian likelihood for
all three models.}
\end{table*}

The updated deuterium abundance therefore weakens the BBN--CMB discrepancy
in $\Lambda$CDM and in the constant-radiation extension, but does not alter
the qualitative model comparison.  Constant $\dNeff$ still increases the helium-4 production
while failing to restore the observed deuterium abundance at the CMB-preferred
baryon density.  vEDE continues to accommodate the CMB value of $\wb$ by
increasing the expansion rate specifically during the residual
deuterium-burning epoch, with only a small change in $Y_p$.
We obtain $\Delta\mathrm{AIC}\simeq+1.8$ for the one-parameter
$\dNeff$ extension and $\Delta\mathrm{AIC}\simeq-3.0$ for the two-parameter
vEDE model. vEDE remains positively selected---it removes the tension
($Q_{\rm DMAP}\simeq0.7\sigma$) and its two extra parameters are justified by the
improved fit against the EDE CMB baryon density---whereas the radiation extension
is not.

\section{Results with $\Lambda$CDM CMB baryon-density priors}
\label{app:lcdmprior}

\begin{figure*}
 \includegraphics[width=\textwidth]{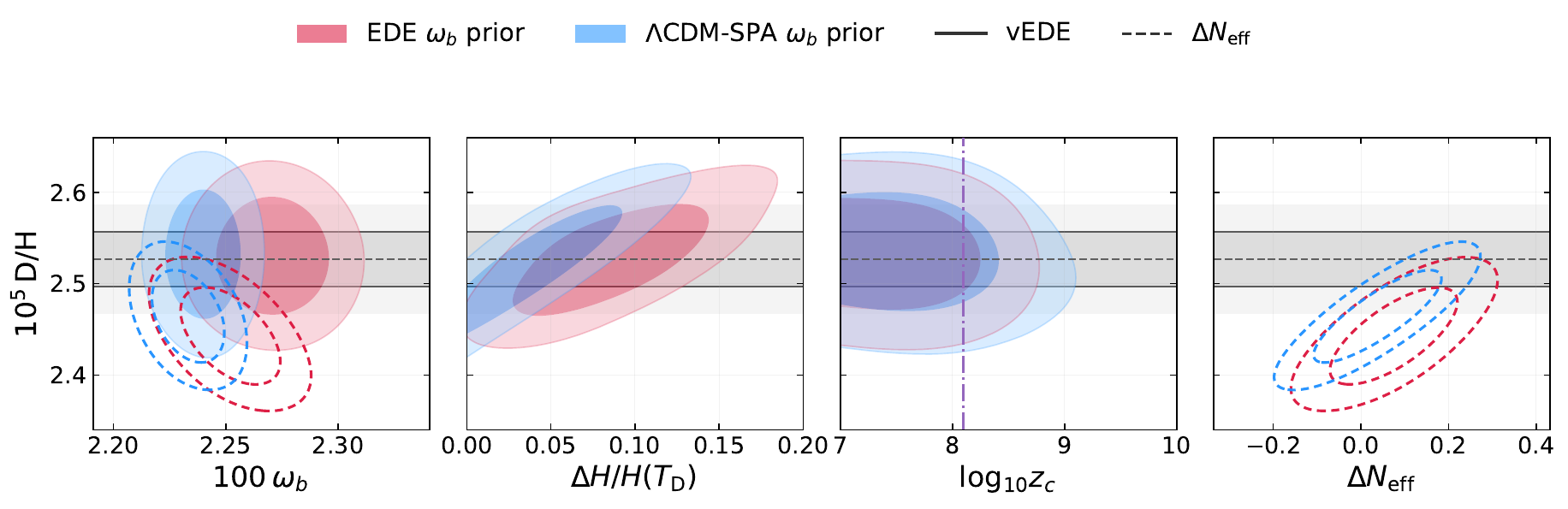}
 \caption{\label{fig:lcdmpriorcomparison}%
 Joint BBN+CMB posterior distributions of D/H against the
 parameters that control the two extensions, using the baseline deuterium
 determination and comparing the main EDE $\omega_b$ prior (red) with the
 $\Lambda$CDM-SPA $\omega_b$ prior (blue).  Contours enclose 68\% and 95\% of
 the posterior.  In the common $100\wb$ panel, filled contours denote vEDE and
 dashed contours denote the $\Delta N_{\rm eff}$ extension; the remaining
 panels show the corresponding model-specific parameters.  The horizontal
 dashed line and grey bands show the central value and $1\sigma$ and $2\sigma$
 ranges of the baseline deuterium determination. On the third panel, the dash-dotted purple line corresponds to $z_\mathrm{D}$ [see after Eq.~\eqref{eq:TD}].}
\end{figure*}

The main analysis uses the baryon density inferred in the EDE
cosmology as the CMB target for all three BBN models.  This choice is
deliberate: we are asking how much non-standard BBN-era expansion would be
needed if the high baryon density characteristic of an early-time resolution
of the Hubble tension were realized.  It is nevertheless useful to repeat the
compatibility estimate with the lower baryon densities inferred in
$\Lambda$CDM-only fits.  We consider both the SPA-only value and the value obtained
after including DESI BAO data~\cite{Camphuis:2025spt},
\begin{align}
 (100\,\wb)^{\Lambda{\rm CDM}}_{\rm SPA}
   &=2.2398\pm0.0095,
 \label{eq:lcdmspa_prior}\\
 (100\,\wb)^{\Lambda{\rm CDM}}_{\rm SPA+DESI}
   &=2.2478\pm0.0091.
 \label{eq:lcdmdesi_prior}
\end{align}
We keep the baseline deuterium abundance,
$10^5\,{\rm D/H}=2.527\pm0.030$, and use the same BBN profile likelihoods.

Results are reported in Table~\ref{tab:lcdmprior}. As expected, lowering the CMB baryon-density target reduces the
baseline discrepancy relative to the main EDE-target analysis.  With the
SPA-only $\Lambda$CDM prior, the $\Lambda$CDM consistency metric is
$Q_{\rm DMAP}\simeq2.1\sigma$, and vEDE removes the discrepancy
($Q_{\rm DMAP}\simeq0.5\sigma$) but is nearly neutral in AIC once its two
extra parameters are counted.  Including DESI BAO shifts the $\Lambda$CDM CMB
baryon density upward, increasing the baseline tension to
$Q_{\rm DMAP}\simeq2.4\sigma$ and giving a mild AIC preference for vEDE,
$\Delta{\rm AIC}\simeq-1.5$.  The $\Delta N_{\rm eff}$ extension does not
improve the compatibility for either prior.  Thus, even within $\Lambda$CDM,
the DESI-induced upward shift moves the comparison toward the high-$\omega_b$
case emphasized in the main text, while the statistical motivation is
strongest for the EDE CMB target.

\begin{table}[!ht]
\caption{\label{tab:lcdmprior}Profile-based consistency with the
$\Lambda$CDM CMB baryon-density priors in
Eqs.~\eqref{eq:lcdmspa_prior} and~\eqref{eq:lcdmdesi_prior}, using the
baseline deuterium abundance.  For each CMB prior, the AIC difference is
referenced to $\Lambda$CDM with the same prior.}
\begin{ruledtabular}
\begin{tabular}{lcccc}
& \multicolumn{2}{c}{SPA} &
\multicolumn{2}{c}{SPA+DESI}\\
Model &
$Q_{\rm DMAP}$ & $\Delta{\rm AIC}$ &
$Q_{\rm DMAP}$ & $\Delta{\rm AIC}$\\
\hline
$\Lambda$CDM &
$2.06\sigma$ & 0 &
$2.41\sigma$ & 0\\
$\Delta N_{\rm eff}$ &
$2.12\sigma$ & $+1.8$ &
$2.41\sigma$ & $+1.5$\\
vEDE &
$0.50\sigma$ & $+0.02$ &
$0.55\sigma$ & $-1.5$\\
\end{tabular}
\end{ruledtabular}
\end{table}

The corresponding posterior distributions are shown in
Fig.~\ref{fig:lcdmpriorcomparison}.  The lower $\Lambda$CDM-SPA baryon-density
prior shifts the vEDE posterior toward a smaller expansion-rate enhancement
than in the main EDE-prior analysis.  For the vEDE run with the
$\Lambda$CDM-SPA prior, we find
\begin{equation}
 \dD = \dH(\TD)=0.050^{+0.020}_{-0.039}
 \label{eq:lcdmprior_dh}
\end{equation}
at 68\% credibility.  This should be compared with
Eq.~\eqref{eq:edepost}: a modest transient enhancement is sufficient for the
lower $\Lambda$CDM-SPA target, while the larger EDE baryon density requires a
larger $\dH(\TD)$.

\section{vEDE posterior in the $f_{\rm vEDE}(z_c)$ parameterization}
\label{app:fvEDE}

As noted in Sec.~\ref{sec:results}, the sampled amplitude
$\dD=\Delta H/H(\TD)$ and the fractional vEDE density $\fede(\zc)$ are two
normalizations of the same component, related through
Eq.~\eqref{eq:mapping}. In this appendix we make explicit how the vEDE
posterior looks when expressed in terms of $\fede(\zc)$ and
$R_{\rm vEDE}(\zc)$ rather than the adopted amplitude $\dD$, using the baseline analysis with
$10^5\,{\rm D/H}=2.527\pm0.030$.

\begin{figure} 
\includegraphics[width=\columnwidth]{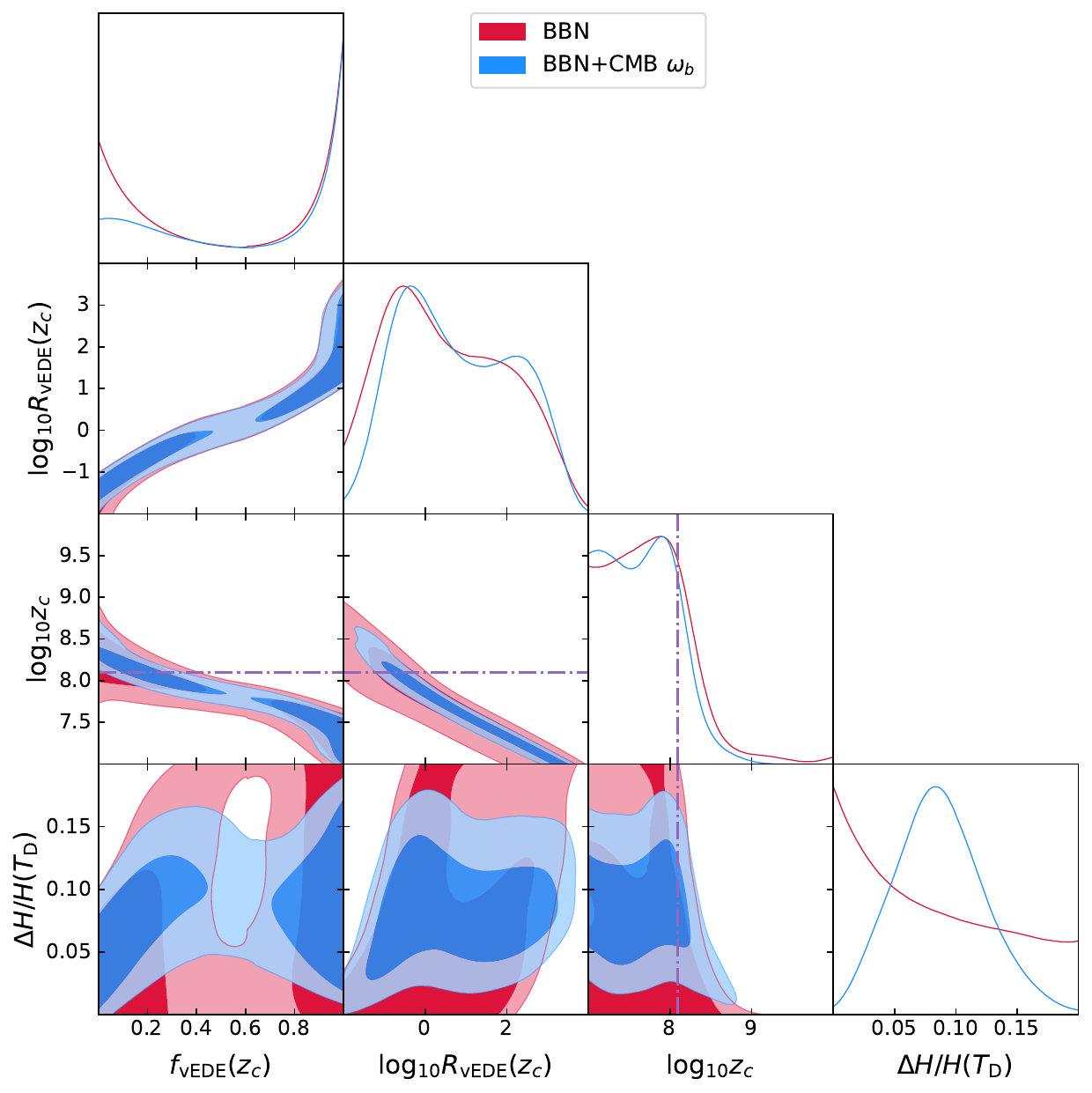}
 \caption{Marginalized posteriors of
 $\{\fede(\zc),\log_{10}R_{\rm vEDE}(\zc),\log_{10}\zc,\dD\}$ in the vEDE model, for the baseline
 deuterium abundance $10^5\,{\rm D/H}=2.527\pm0.030$. Red contours show the
 BBN-only analysis and blue contours the combined BBN+CMB $\wb$ analysis;
 dark and light shadings contain 68\% and 95\% of the posterior probability.
 Here $R_{\rm vEDE}(\zc)\equiv \rho_\phi(\zc)/[\rho_{\rm tot}(\zc)-\rho_\phi(\zc)]
 = \fede(\zc)/[1-\fede(\zc)]$.
 While the expansion-rate amplitude $\dD$ is well constrained once the CMB
 $\wb$ prior is included, the fractional density $\fede(\zc)$ remains
 essentially unconstrained across its prior range and the critical redshift is
 bounded only from above ($\log_{10}\zc<8.3$ at $95\%$). \label{fig:fvEDEtriangle}}
\end{figure}

The result is shown in Fig.~\ref{fig:fvEDEtriangle}. As anticipated,
the fractional density $\fede(\zc)$ is essentially unconstrained, spreading over
almost its entire prior range, while the critical redshift is
bounded only from above: the combined BBN+CMB analysis yields the one-sided
limits $\log_{10}\zc<8.3$ ($95\%$), with the marginal
posterior flat toward lower redshifts down to the prior boundary. Physically, a
broad family of critical redshifts and fractional amplitudes reproduces the same
expansion-rate enhancement near deuterium freeze-out and hence the same
light-element abundances; only transitions substantially earlier than the
deuterium-burning epoch, whose contribution is diluted away before D/H freezes
out, are disfavoured, which produces the upper limit on $\zc$. By contrast, the
derived amplitude $\dD$ is well localized in the combined BBN+CMB analysis,
consistent with Eq.~\eqref{eq:edepost}. This confirms that $\dD$, and not
$\fede(\zc)$ or $\zc$, is the physically meaningful vEDE amplitude at the BBN
epoch, and justifies its use as the sampled amplitude parameter in the main
analysis.

The apparent bimodality of the $\fede(\zc)$ posterior, with support
piling up near $\fede\simeq0$ and $\fede\simeq1$, is a prior-volume effect of
the nonlinear map between the sampled pair $(\dD,\log_{10}\zc)$ and the derived
$\fede(\zc)$; the posterior in the sampled parameters is smooth and unimodal.
At fixed amplitude $\dD$, Eqs.~\eqref{eq:fede} and~\eqref{eq:mapping} give an
$\fede(\zc)$ that is \emph{minimized} when the transition coincides with the
pivot, $\zc\sim z_{\rm D}$, and rises toward unity as $\zc$ moves away from
$z_{\rm D}$ in either direction. For $\zc\gg z_{\rm D}$ the component must have
been large at the transition to survive its post-transition dilution
[$\propto(1+z)^{3(1+w_n)}$, faster than radiation] and still supply the
required $\dD$ at $\TD$; for $\zc\ll z_{\rm D}$ the frozen plateau density
fixed by $\dD$ is compared to a radiation density that has since dropped as
$(1+z)^4$, so its fractional contribution at $\zc$ again approaches one.
The pile-up at $\fede\simeq1$ therefore collects the large prior volume of
critical redshifts far from $z_{\rm D}$ across the wide flat range
$\log_{10}\zc\in[7,10]$, while the pile-up at $\fede\simeq0$ is simply the
no-vEDE region $\dD\to0$, which the BBN-only likelihood does not exclude.
Intermediate values $\fede(\zc)\sim0.5$ require $\dD>0$ \emph{and}
$\zc\sim z_{\rm D}$ simultaneously, a thin slice of the prior that is
correspondingly depleted. This is precisely the ambiguity that motivates
sampling $\dD$ rather than $\fede$.

It is useful to translate the vEDE amplitude into the variable
\begin{equation}
  R_{\rm vEDE}(z_c)\equiv
  \frac{\rho_{\rm vEDE}(z_c)}{\rho_{\rm tot}(z_c)-\rho_{\rm vEDE}(z_c)}
  =\frac{f_{\rm vEDE}(z_c)}{1-f_{\rm vEDE}(z_c)} \, .
\end{equation}
This is the ratio used in Ref.~\cite{Sobotka:2024tat} to characterize
very-early scalar-field episodes.  Since $R_{\rm vEDE}$ is unbounded above,
large values correspond to histories in which the scalar field temporarily
dominates the energy density near $z_c$.
The BBN likelihood constrains primarily the local expansion-rate
enhancement during deuterium burning, $\Delta H/H(T_D)$, rather than
$R_{\rm vEDE}(z_c)$ itself. 
Consequently, sizeable values of $R_{\rm vEDE}(z_c)$
remain allowed whenever the scalar-field episode is arranged so that the
expansion rate around $T_D$ gives the required deuterium abundance while
leaving helium production nearly unchanged.  
In this sense, BBN provides a
complementary constraint to probes of the post-BBN evolution: it constrains the scalar field through its effect on the expansion rate
during nucleosynthesis, while allowing the peak scalar fraction at $z_c$
to be much larger.

\bibliographystyle{apsrev4-2}
\bibliography{biblio}

\end{document}